\documentclass[a4paper]{article}
\usepackage{algorithm}
\usepackage{algorithmic}
\usepackage{INTERSPEECH2021}
\usepackage{amsfonts,subfigure,cite,epsfig,etoolbox,enumitem}
\usepackage{multirow}

\title{SpecAugment++: A Hidden Space Data Augmentation Method for Acoustic Scene Classification}
\name{Helin Wang$^1$, Yuexian Zou$^{1,2,*}$, Wenwu Wang$^3$ \thanks{*Corresponding author: zouyx@pku.edu.cn}
\thanks{This paper was partially supported by the Shenzhen Science \& Technology Fundamental Research Programs 
(No: JCYJ20180507182908274 \& JSGG20191129105421211) 
and GXWD20201231165807007-20200814115301001.
}}
\address{
  $^1$ADSPLAB, School of ECE, Peking University, Shenzhen, China\\
  $^2$Peng Cheng Laboratory, Shenzhen, China\\
  $^3$Center for Vision, Speech and Signal Processing, University of Surrey, UK}
\email{wanghl15@pku.edu.cn, zouyx@pku.edu.cn, w.wang@surrey.ac.uk}

\begin{document}

\maketitle
\begin{abstract}
In this paper, 
we present SpecAugment++, 
a novel data augmentation method for deep neural networks based acoustic scene classification (ASC).
Different from other popular data augmentation methods such as SpecAugment and mixup that only work on the input space,
SpecAugment++ is applied to both the input space and the hidden space of the deep neural networks to enhance the input and the intermediate feature representations.
For an intermediate hidden state, 
the augmentation techniques consist of masking blocks of frequency channels
and masking blocks of time frames,
which improve generalization by enabling a model to attend not only to the most discriminative parts of the feature,  
but also the entire parts.
Apart from using zeros for masking, 
we also examine two approaches for masking based on the use of other samples within the mini-batch, 
which helps introduce noises to the networks to make them more discriminative
for classification.
The experimental results on the DCASE 2018 Task1 dataset and DCASE 2019 Task1 dataset show that 
our proposed method can obtain $3.6\%$ and $4.7\%$ accuracy gains over a strong baseline without augmentation (\textit{i.e.} CP-ResNet) respectively, 
and outperforms other previous data augmentation methods.

\end{abstract}

\section{Introduction}
Deep learning has been successfully applied to various problems in Detection and Classification of Acoustic Scenes and Events (DCASE) \cite{kong2016deep,valenti2016dcase,mesaros2019sound},
where convolutional neural networks (CNNs) \cite{zhang2015robust,Wang2020,wang2020modeling}, recurrent neural networks (RNNs) \cite{parascandolo2016recurrent} and convolutional recurrent neural networks (CRNNs) \cite{cakir2017convolutional} are often used as the network architectures.
However, due to the lack of training data, these models are prone to overfitting \cite{zou2019improved}.

Data augmentation methods have been widely used to overcome the problem of the limited data in the DCASE community,
including the waveform-based and spectrogram-based data augmentation methods.
Among the waveform-based methods, cropping is one of the common and effective approaches \cite{aytar2016soundnet,tokozume2017learning,piczak2015environmental}.
Salamon and Bello \cite{salamon2017deep} proposed the usage of additional training data generated by time stretching, pitch shifting, dynamic range compression, and adding background noise chosen from an external dataset,
which are also applied to the raw waveform.
In \cite{park2019specaugment}, Park \textit{et al.} proposed SpecAugment, 
an augmentation method that operates on the log mel spectrogram of the input audio
and achieved the state-of-the-art performance for automatic speech recognition (ASR).
SpecAugment consists of three kinds of deformations of the log mel spectrogram (\textit{i.e.} time warping, time masking and frequency masking)
to generate difficult-recognition samples,
and converts ASR from an over-fitting to an under-fitting problem.
However, these augmentation methods are only applied to the input of the deep neural networks,
while augmenting the hidden space is not studied.
Mixup \cite{zhang2017mixup,wang2021global} and Between-Class (BC) learning \cite{tokozume2018learning} are also popular data augmentation methods for the DCASE tasks,
which generate new data samples by mixing multiple audio samples and design the learning method for training a model to output the prediction of the mixed samples.
SpeechMix \cite{jindal2020speechmix} is a generalization of BC learning and it interpolates latent representations of hidden states.
However, it is hard to determaine the label of the mixed audio as the energy and the distribution of the raw audio is quite different.
In addition, Chen \textit{et al.} \cite{chen2019integrating} utilized the auxiliary classifier GAN (ACGAN) to generate fake samples for data augmentation,
however, an extra discriminator is needed, as a result,
the convergence of the networks becomes more difficult.

In this paper, we propose a novel data augmentation method (\textit{i.e.} SpecAugment++),
which is applied to both the input spectrograms and the hidden states of the deep neural networks.
Inspired by the SpecAugment \cite{park2019specaugment},
two kinds of deformations of the hidden states (\textit{e.g.} the intermediate feature maps in CNN) are employed,
containing masking blocks of frequency channels and masking blocks of time frames.
These methods improve the generalization ability of a model by allowing it to attend not only to the most discriminative
time frames and frequency channels of audio, 
but also to the entire temporal frames and frequency channels.
Time warping is not applied because it contributes little to the classification \cite{park2019specaugment,wang2020environmental}.
Apart from exploiting the augmentation for the hidden space,
we study the influence of three schemes for time masking and frequency masking.
The first is zero-value masking (ZM), which directly masks consecutive time frames and frequency channels with zeros.
The other two schemes are the mini-batch based mixture masking (MM) and the mini-batch based cutting masking (CM),
which utilize the time frames and frequency channels of other samples within the mini-batch for masking.
These methods can be seen as introducing additional noises generated from the dataset,
and guide the networks to be more discriminative for classification.
Different from the mixup \cite{zhang2017mixup} and BC learning \cite{tokozume2018learning},
the labels of the augmented data are not changed
so that the training is more stable.
The proposed method is simple and computationally cheap to apply,
which is evaluated on two benchmark acoustic scene classification (ASC) datasets
(DCASE 2018 \cite{mesaros2018multi} and DCASE 2019 \cite{mesaros2019acoustic} ASC task 1A datasets)
and outperforms the state-of-the-art data augmentation methods.

\section{Proposed Method}
In this section, we introduce the proposed augmentation method,
which is constructed to act on either the input spectrograms or the intermediate hidden states of the neural networks.
Time masking and frequency masking are employed,
and the motivation is to promote the networks to be robust to deformations in the partial loss of the temporal information and the partial loss of the frequency information in each layer.
Three types of masking schemes are developed,
aiming to make better use of the data within the mini-batch.
The augmentation techniques and the masking schemes are detailed as follows.

\begin{figure}[t] 
  \centering
  \subfigure[The target sample.]{\label{f1}
  \includegraphics[width=3in]{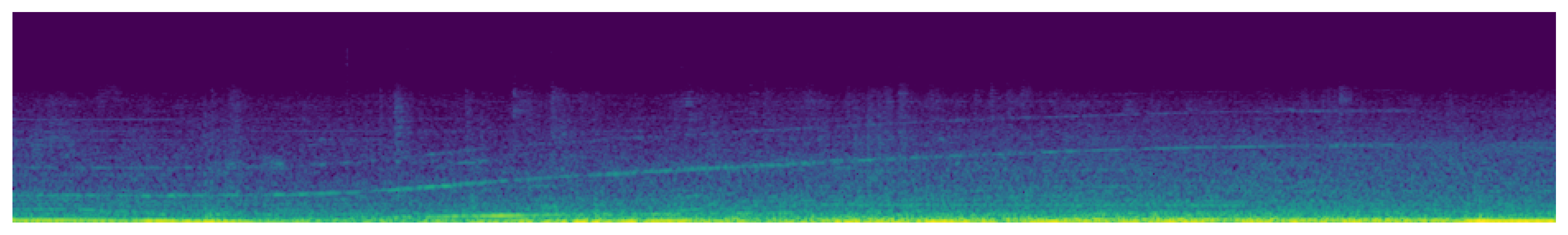}}
  \subfigure[Another sample within the same mini-batch.]{\label{f2}
  \includegraphics[width=3in]{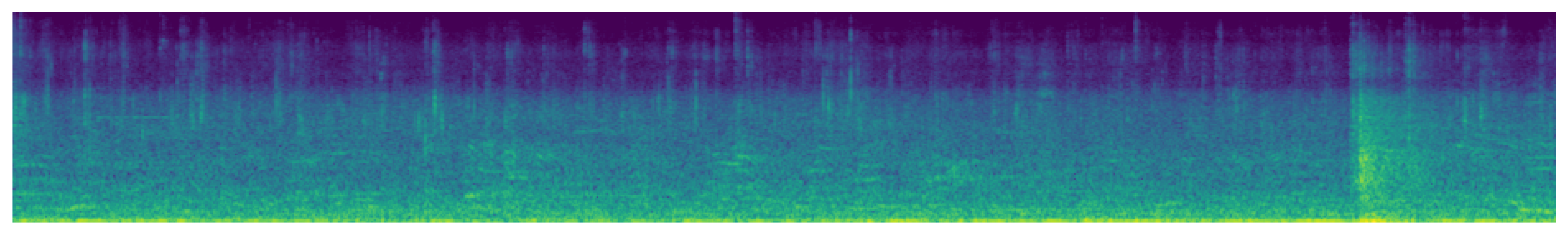}}
  \subfigure[The augmented sample by ZM.]{\label{f3}
  \includegraphics[width=3in]{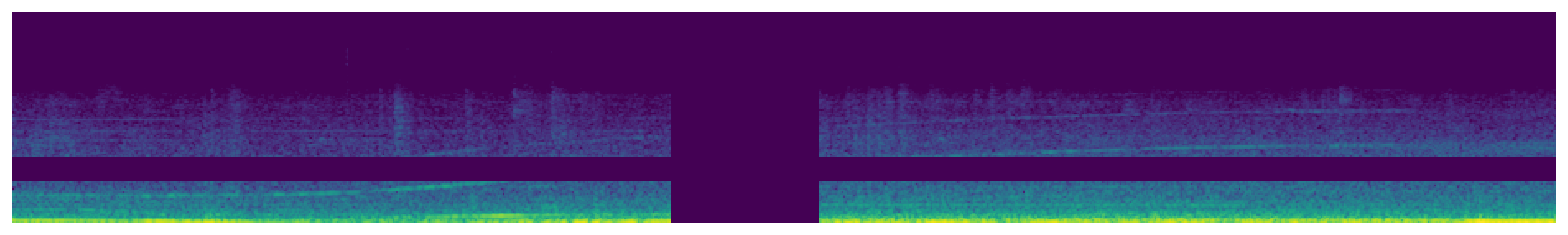}}
  \subfigure[The augmented sample by MM.]{\label{f4}
  \includegraphics[width=3in]{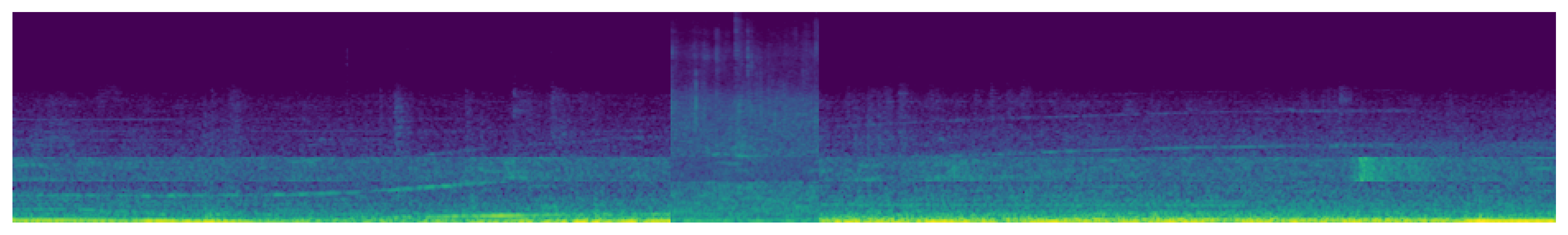}}
  \subfigure[The augmented sample by CM.]{\label{f5}
  \includegraphics[width=3in]{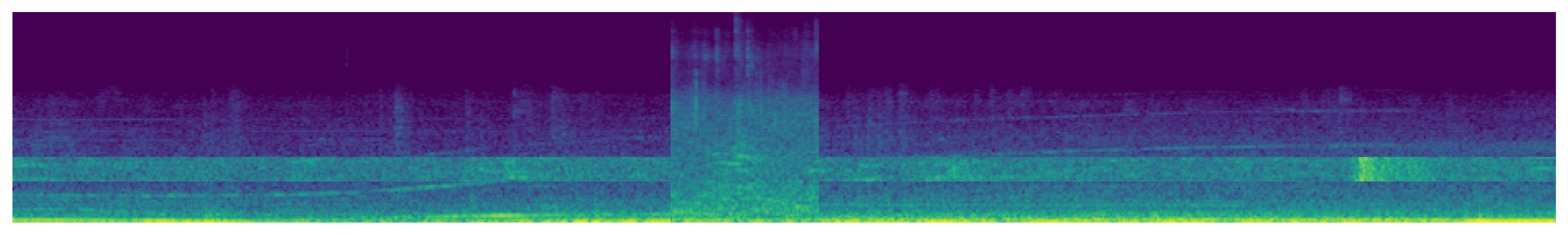}}
  \caption{Augmentations applied to the target sample.
  Here, the input log mel spectrogram is shown as an example,
  similarly to the augmentation applied to the intermediate hidden states.
  From top to bottom, the figures depict the log mel spectrogram of the target sample with no augmentation, 
  another sample within the same mini-batch,
  the zero-value masking (ZM) applied,
  the mini-batch based mixture masking (MM) applied
  and the mini-batch based cutting masking (CM) applied.
  }
  \label{fig1}
\end{figure}

\subsection{Augmentation Techniques}
Let $x \in \mathbb{R}^{T \times F}$ denote the intermediate hidden state (or the input spectrogram),
where $T$ and $F$ denote the number of time frames and frequency channels, respectively.
Time masking is applied so that $t$ consecutive time frames $\left[t_{0}, t_{0}+t\right]$ are masked (which means replacing the elements by zeros or other values), 
where $t$ is chosen from a uniform distribution from $0$ to the time mask parameter $t^{\prime}$, 
and $t_{0}$ is chosen from $\left[0, T-t\right]$.
Similarly, frequency masking is applied so that $f$ consecutive frequency channels $\left[f_{0}, f_{0}+f\right]$ are masked, 
where $f$ is first chosen from a uniform distribution from $0$ to the frequency mask parameter $f^{\prime}$,
and $f_{0}$ is chosen from $\left[0, F-f\right]$.
To simplify, the same time masking and frequency masking (\textit{i.e.} the same $t_{0}$, $t$, $f_{0}$ and $f$) are applied to each intermediate hidden state in the same layer for a training sample.
For instance, if there are $C$ feature maps in the $l$-th layer of a CNN model,
these $C$ feature maps share the same temporal and frequency regions for masking of a single sample during one training iteration.

\begin{algorithm}[t]
  \caption{Zero-value masking (ZM)}
  \label{alg1}
  \begin{algorithmic}[1]
    \REQUIRE The hidden state of the target sample $x \in \mathbb{R}^{T \times F}$;
    The number of the consecutive time frames $t$; The starting time index $t_{0}$;
    The number of the consecutive frequency channels $f$; The starting frequency index $f_{0}$;
    \ENSURE The augmented hidden state of the target sample $x^{\prime} \in \mathbb{R}^{T \times F}$;
    \STATE $x^{\prime}=x$;
    \FOR{$i = t_{0}, \ldots, t_{0}+t$}
      \FOR{$j = 0, \ldots, F$}
        \STATE $x^{\prime}\left[i,j\right]=0$;
      \ENDFOR
    \ENDFOR
    \FOR{$i = 0, \ldots, T$}
      \FOR{$j = f_{0}, \ldots, f_{0}+f$}
        \STATE $x^{\prime}\left[i,j\right]=0$;
      \ENDFOR
    \ENDFOR
    \RETURN $x^{\prime}$;
  \end{algorithmic}
\end{algorithm}

\begin{algorithm}[t]
  \caption{Mini-batch based mixture masking (MM)}
  \label{alg1}
  \begin{algorithmic}[1]
    \REQUIRE The hidden state of the target sample $x \in \mathbb{R}^{T \times F}$;
    The hidden state of another sample within the same mini-batch $y \in \mathbb{R}^{T \times F}$;
    The number of the consecutive time frames $t$; The starting time index $t_{0}$;
    The number of the consecutive frequency channels $f$; The starting frequency index $f_{0}$;
    \ENSURE The augmented hidden state of the target sample $x^{\prime} \in \mathbb{R}^{T \times F}$;
    \STATE $x^{\prime}=x$;
    \FOR{$i = t_{0}, \ldots, t_{0}+t$}
      \FOR{$j = 0, \ldots, F$}
        \STATE $x^{\prime}\left[i,j\right]=\frac{1}{2}\left(x[i,j]+y[i,j]\right)$;
      \ENDFOR
    \ENDFOR
    \FOR{$i = 0, \ldots, T$}
      \FOR{$j = f_{0}, \ldots, f_{0}+f$}
        \STATE $x^{\prime}\left[i,j\right]=\frac{1}{2}\left(x[i,j]+y[i,j]\right)$;
      \ENDFOR
    \ENDFOR
    \RETURN $x^{\prime}$;
  \end{algorithmic}
\end{algorithm}

\subsection{Masking Schemes}
As shown in Figure~\ref{fig1}, we present three masking schemes, including the zero-value masking (ZM), the mini-batch based mixture masking (MM) and the mini-batch based cutting masking (CM).
ZM directly applies the zero value for the masking regions of the target sample.
MM and CM utilize the time frames and frequency channels from another sample for masking.
To explain,
if the hidden state in the $l$-th layer of the target sample is to be augmented,
we randomly select another sample within the same mini-batch as the target sample
and use the hidden state in the $l$-th layer of the selected sample for masking.
The difference is that MM mixes the masking regions of the hidden states of the two samples by the mean, 
while CM fills the masking regions of the target sample with the same regions of the selected sample.
The details are summarized in Algorithm 1, Algorithm 2 and Algorithm 3, respectively.
We further discuss their pros and cons in Section 4.3.

\begin{algorithm}[t]
  \caption{Mini-batch based cutting masking (CM)}
  \label{alg1}
  \begin{algorithmic}[1]
    \REQUIRE The hidden state of the target sample $x \in \mathbb{R}^{T \times F}$;
    The hidden state of another sample within the same mini-batch $y \in \mathbb{R}^{T \times F}$;
    The number of the consecutive time frames $t$; The starting time index $t_{0}$;
    The number of the consecutive frequency channels $f$; The starting frequency index $f_{0}$;
    \ENSURE The augmented hidden state of the target sample $x^{\prime} \in \mathbb{R}^{T \times F}$;
    \STATE $x^{\prime}=x$;
    \FOR{$i = t_{0}, \ldots, t_{0}+t$}
      \FOR{$j = 0, \ldots, F$}
        \STATE $x^{\prime}\left[i,j\right]=y[i,j]$;
      \ENDFOR
    \ENDFOR
    \FOR{$i = 0, \ldots, T$}
      \FOR{$j = f_{0}, \ldots, f_{0}+f$}
        \STATE $x^{\prime}\left[i,j\right]=y[i,j]$;
      \ENDFOR
    \ENDFOR
    \RETURN $x^{\prime}$;
  \end{algorithmic}
\end{algorithm}

\section{Models}
We use CP-ResNet\footnote{https://github.com/kkoutini/cpjku\_dcase19/} \cite{koutini2019receptive} as the base model for the ASC task.
CP-ResNet is a variant of ResNet \cite{he2016deep} by adapting the audio tasks using receptive field (RF) regularization,
which shows the best performance among single models for ASC \cite{koutini2020low}.
The network architecture is detailed in Table~\ref{t1}.
There are a series of residual convolutional blocks with a kernel size of $3\times3$ or $1\times1$,
followed by a global average pooling function and a fully-connected layer.
CP-ResNet has a RF of $87\times87$ and a total of $6,053,780$ trainable parameters.

Our proposed method is applied to the CP-ResNet between the model blocks.
More specifically, the augmentation is applied to the input log mel spectrograms (denoted as Layer $0$),
the hidden states before residual block (RB) $1$ (denoted as Layer $1$), before RB $5$ (denoted as Layer $2$), before RB $9$ (denoted as Layer $3$) and after RB $12$ (denoted as Layer $4$), respectively.
Hence, both the input space (Layer $0$) and the hidden space (Layers $1$-$4$) can be enhanced.

\begin{table}[t]
  \begin{center}
  \caption{Network architecture of CP-ResNet} \label{t1}
  \begin{tabular}{|c|c|}
      \hline
      \textbf{RB Number} & \textbf{RB Config}\\
      \hline
       & $5 \times 5 $, C=$64$, S=$2$\\
      \hline
      1 & $3 \times 3 $, $1 \times 1 $, C=$128$, S=$1$, P\\
      2 & $3 \times 3 $, $3 \times 3 $, C=$128$, S=$1$, P\\
      3 & $3 \times 3 $, $3 \times 3 $, C=$128$, S=$1$\\
      4 & $3 \times 3 $, $3 \times 3 $, C=$128$, S=$1$, P\\
      \hline
      5 & $3 \times 3 $, $1 \times 1 $, C=$256$, S=$1$\\
      6 & $1 \times 1 $, $1 \times 1 $, C=$256$, S=$1$\\
      7 & $1 \times 1 $, $1 \times 1 $, C=$256$, S=$1$\\
      8 & $1 \times 1 $, $1 \times 1 $, C=$256$, S=$1$\\
      \hline
      9 & $1 \times 1 $, $1 \times 1 $, C=$512$, S=$1$\\
      10 & $1 \times 1 $, $1 \times 1 $, C=$512$, S=$1$\\
      11 & $1 \times 1 $, $1 \times 1 $, C=$512$, S=$1$\\
      12 & $1 \times 1 $, $1 \times 1 $, C=$512$, S=$1$\\
      \hline      
      \multicolumn{2}{p{200pt}}{RB: Residual Block; S: stride; P: $2\times2$ max pooling after the block;
      C: Number of channels.}\\
\end{tabular}
\label{t1}
\end{center}
\end{table}

\section{Experiments}

\subsection{Datasets}
The experiments were conducted on the DCASE 2018 \cite{mesaros2018multi} and DCASE 2019 \cite{mesaros2019acoustic} ASC task 1a datasets,
which are commonly used benchmark datasets for ASC.
The DCASE 2018 task 1a dataset \cite{mesaros2018multi} consists of about $17$ hours of audio for training ($6122$ $10$-second clips)
and $7$ hours for evaluation ($2518$ $10$-second clips).
The DCASE 2019 task 1a dataset \cite{mesaros2019acoustic} contains a total of $40$ hours of audio
($9185$ clips in the training set, $4185$ clips in the test set).
Each recording belongs to one out of $10$ possible classes.

\subsection{Setups}
For all experiments, we follow the same settings as the state of the art \cite{koutini2019receptive,sakashita2018acoustic}
to ensure a fair comparison.
Specifically,
the input is down-sampled to $22.05$kHz and applied a Short Time Fourier Transform (STFT)
with a window size of $2048$ and $25\%$ overlap,
followed by a Mel-scaled filter bank on perceptually weighted spectrograms.
This results in $256$ Mel frequency bins and around $43$ frames per second. 
The input frames are normalized to zero-mean and unit variance according to the training set.
The Adam optimizer \cite{kingma2014adam} is used for a total of $350$ epochs, 
with an initial learning rate of $1 \times 10^{-4}$.
The learning rate decays linearly from epoch $50$ until $250$ where it reaches $5 \times 10^{-6}$.
Then we train for another $100$ epochs with the minimum learning rate $5 \times 10^{-6}$.
The models are evaluated on the test set after $350$ epochs of training. 
Each experiment is repeated $3$ times,
and we report the mean and the standard deviation of these runs.
When using SpecAugment++\footnote{https://github.com/WangHelin1997/SpecAugment-plus}, 
we randomly select a layer to perform the augmentation from a set of layers (Layers $0$-$4$).
Unless otherwise stated, 
we set the hyperparameters $t^{\prime}$ as $43$ and $f^{\prime}$ as $26$ in our experiments.
Thus, around $10\%$ of the time frames and frequency channels are masked.

\begin{table}[t]
  \begin{center}
  \caption{Comparisons of classification accuracy on different data dugmentation methods ($\%$).} \label{t2}    
  \begin{tabular}{|c|c|c|}
      \hline
      \textbf{Augmentation Method} & \textbf{DCASE 18} & \textbf{DCASE 19}\\
      \hline
      No augmentation \cite{koutini2019receptive} & $74.3\pm0.59$ & $78.9\pm0.80$ \\
      mixup (2017) \cite{zhang2017mixup} & $75.5\pm0.62$ & $79.3\pm0.71$ \\
      SpecAugment (2019) \cite{park2019specaugment} & $74.9\pm0.81$ & $79.1\pm1.05$ \\
      BC Learning (2017) \cite{tokozume2018learning} & $75.8\pm0.66$ & $80.0\pm0.76$ \\
      SpeechMix (2020) \cite{jindal2020speechmix} & $75.8\pm0.48$ & $80.7\pm0.69$ \\
      \hline
      \textbf{SpecAugment++ (ZM)} & $76.2\pm0.59$ & $80.6\pm0.82$ \\
      \textbf{SpecAugment++ (MM)} & $\bf{77.0\pm0.52}$ & $\bf{82.6\pm0.66}$ \\
      \textbf{SpecAugment++ (CM)} & $76.9\pm0.73$ & $81.4\pm0.94$ \\
      \hline
\end{tabular}
\label{t2}
\end{center}
\end{table}

\subsection{Results}

We compare the performance of our proposed method with the state of the art \cite{zhang2017mixup,park2019specaugment,tokozume2018learning,jindal2020speechmix},
and the results are summarized in Table~\ref{t2}.
Under a strong baseline model \cite{koutini2019receptive}, the proposed SpecAugment++ significantly improves the performance on both datasets,
and outperforms the other previous data augmentation methods (\textit{i.e.} mixup \cite{zhang2017mixup}, SpecAugment \cite{park2019specaugment}, BC Learning \cite{tokozume2018learning} and SpeechMix \cite{jindal2020speechmix}).
Among them, SpecAugment++ with MM achieves the highest accuracy gain ($3.6\%$ on DCASE 18 and $4.7\%$ on DCASE 19),
followed by SpecAugment++ with CM ($3.5\%$ on DCASE 18 and $3.2\%$ on DCASE 19) and SpecAugment++ with ZM ($2.6\%$ on DCASE 18 and $2.2\%$ on DCASE 19).
These feature masking schemes improve generalization and robustness by enabling a model to attend not only to the most discriminative
time frames and frequency channels, but also to the entire temporal and frequency parts.

\begin{figure}[t]
  \centering
  \subfigure[Comparison of different time masking ratios.]{\label{ff1}
  \includegraphics[width=3in]{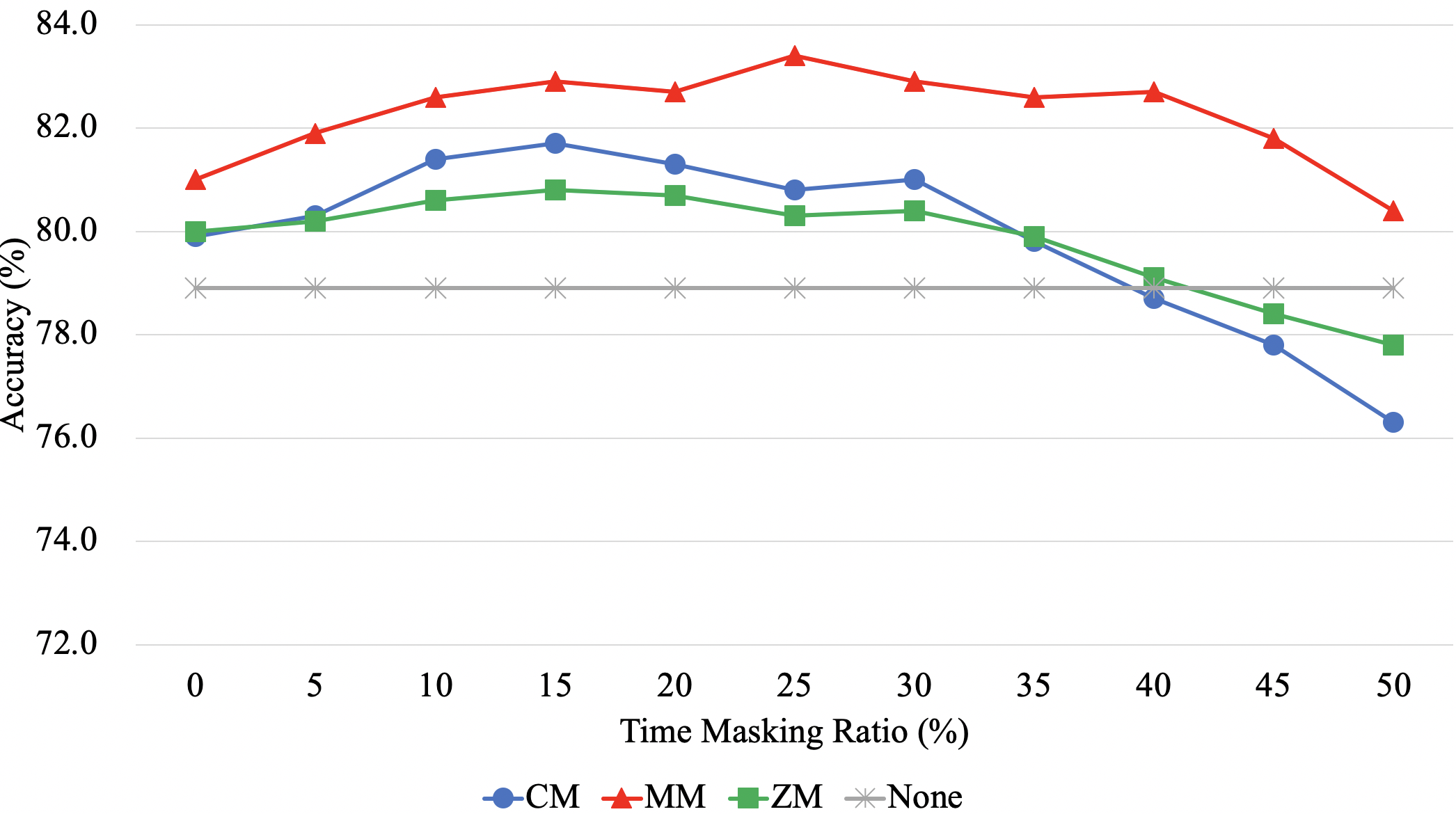}}
  \subfigure[Comparison of difference frequency masking ratios.]{\label{ff2}
  \includegraphics[width=3in]{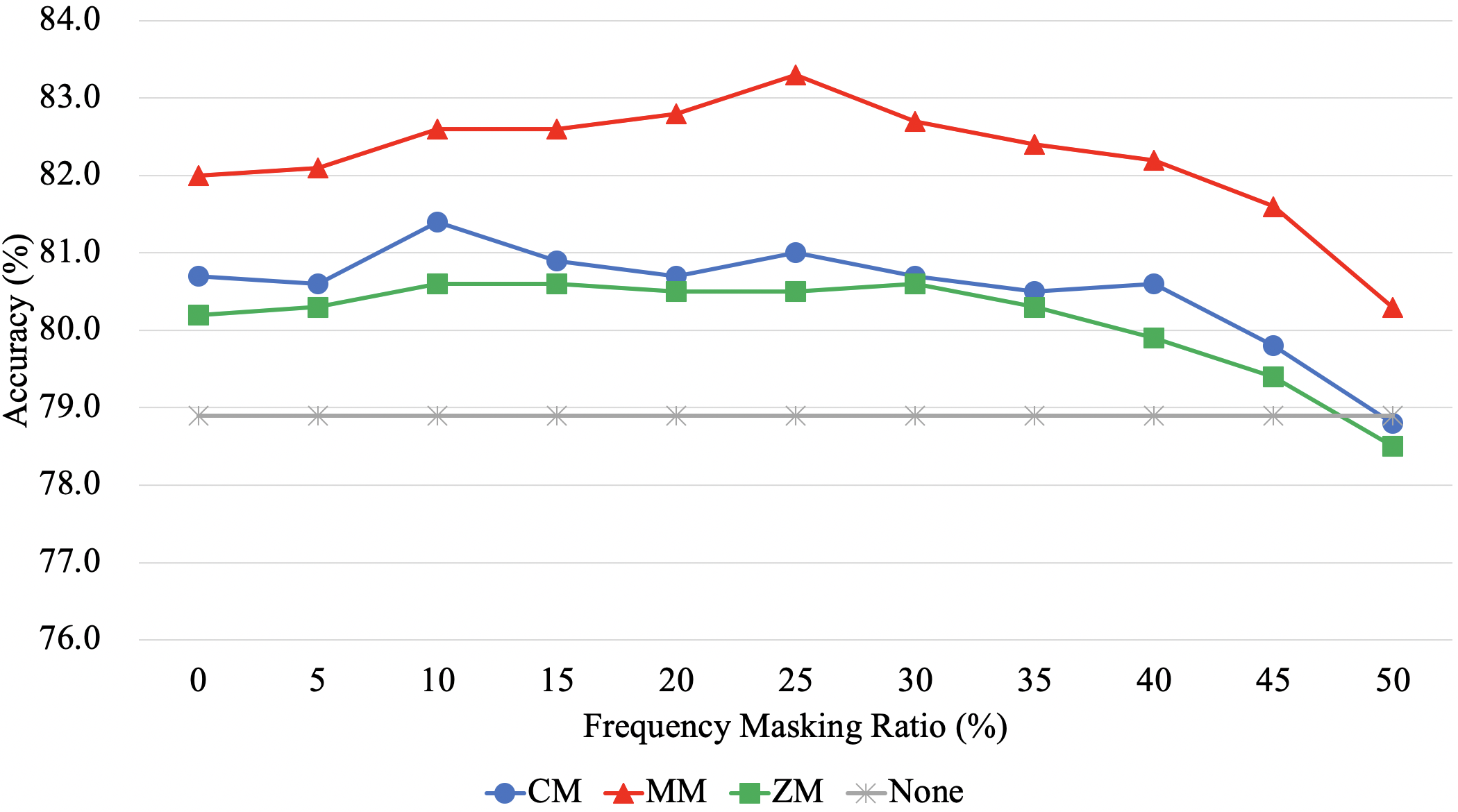}}
  \caption{Accuracy comparison of difference time masking ratios and frequency masking ratios on the DCASE 19.
  Here, we report the mean of three experiments.
  To evaluate the influence of time masking, we keep a frequency masking ratio of $10\%$.
  Similarly, a time masking ratio of $10\%$ is kept when comparing different frequency masking ratios.
  }
  \label{fig2}
\end{figure}

\noindent
\textbf{Comparison of ZM, MM and CM. } SpecAugment++ with ZM can be viewed as a generalization of SpecAugment \cite{park2019specaugment},
where the zero-value masking is applied to both the input log mel spectrograms and the hidden states of the network.
ZM can let a model focus on less discriminative time frames and frequency channels,
while not being efficient enough due to unused parts of the audio.
MM and CM performs better than ZM because the time frames and frequency channels are masked by the corresponding parts of other audio,
which introduce the interference and force the network to be more discriminative.
However, CM may cause unnatural artifacts due to the abruptness of the masking parts and the original hidden states.
In comparison, MM is a more natural choice, 
as the information of the whole target audio sample is retained and the interference of time frames and frequency channels from other audio samples
can enhance the robustness.
In addition, compared with mixup \cite{zhang2017mixup}, BC Learning \cite{tokozume2018learning} and SpeechMix \cite{jindal2020speechmix}
which also mix the audio samples,
ZM and CM do not change the label of the target audio sample so that the training is more stable.

\subsection{Ablation Study}
In order to explore the effectiveness of the time masking and frequency masking,
we compare different time masking ratios and frequency masking ratios
and show the results in Figure~\ref{fig2}.
It can be seen that both time masking
and frequency masking improve the performance.
As the ratios of time masking and frequency masking increase, 
the accuracy of all the three types of masking (\textit{i.e.} ZM, MM and CM) is boosted and then drops, 
which shows the optimal performance when the ratio is $\left[10\%,25\%\right]$.
We argue that a masking ratio of around $10\%$ is enough to
guide a model focus on less discriminative time frames and frequency channels 
and increase the robustness.
However, if the masking ratio is too large (\textit{e.g.} over $40\%$),
the information of the original audio may be lost significantly,
which confuses the model severely and makes the model hard to be trained.

In addition, we investigate different sets of layers for SpecAugment++ on the DCASE 19 dataset,
and results are shown in Table~\ref{t3}.
When no augmentation is performed, the model accuracy is $78.9\%$.
The performance tends to improve with the augmentation both on the input spectrogram (Layer $0$) and on the hidden states (Layers $1$-$4$),
and the improvement is more significant when we apply the SpecAugment++ to the layers near the input.
These layers learn discriminative features such as frequency response which is quite similar to the human perception \cite{tokozume2018learning}.
Our proposed method achieves the best performance when used to all the layers.

\begin{table}[t]
  \begin{center}
  \caption{Ablation analysis of the layers set used for the augmentation.
  While applying the SpecAugment++, 
  we randomly select a layer to perform the time masking and frequency masking
  from the set of layers.
  Layer $0$ denotes the input log mel spectrogram 
  and Layers $1$-$4$ denote the hidden states, which have been described in Section 3.
  Here, we report the mean of the three experiments on the DCASE 19 ($\%$).
  } \label{t2}    
  \begin{tabular}{|c|c|c|c|c|}
      \hline
      \multicolumn{2}{|c|}{} & \textbf{CM} & \textbf{MM}& \textbf{ZM}\\
      \hline
      \multirow{10}{*}{Layers Set} & - & $78.9$ & $78.9$ & $78.9$ \\
      & $\{0\}$ & $80.5$ & $81.8$ & $79.8$ \\
      & $\{1\}$ & $80.1$ & $81.2$ & $79.6$ \\
      & $\{2\}$ & $80.2$ & $81.0$ & $79.4$ \\
      & $\{3\}$ & $79.7$ & $80.6$ & $79.5$ \\
      & $\{4\}$ & $79.9$ & $80.7$ & $79.3$ \\
      & $\{0,1\}$ & $80.9$ & $82.1$ & $80.0$ \\
      & $\{0,1,2\}$ & $81.1$ & $82.4$ & $80.5$ \\
      & $\{0,1,2,3\}$ & $81.2$ & $82.3$ & $80.6$ \\
      & $\{0,1,2,3,4\}$ & $81.4$ & $82.6$ & $80.6$ \\
      \hline
\end{tabular}
\label{t3}
\end{center}
\end{table}

\section{Conclusions}
We have presented a hidden space data augmentation method for acoustic scene classification,
which can broaden the intermediate feature representations for deep neural networks.
Our proposed method is simple and computationally cheap to apply,
which has shown better performance than the state-of-the-art methods.
We will further explore its applications to other tasks in the DCASE community,
such as sound event detection and audio tagging.



\bibliographystyle{IEEEtran}

\bibliography{mybib}


\end{document}